\documentclass[submission,copyright]{eptcs}

\providecommand{\event}{PLACES 2014}
\def\Title{Towards Composable Concurrency Abstractions}
\title{\Title\footnote{An appendix with a complete discussion of results is available at {\scriptsize \url{http://soft.vub.ac.be/\~jswalens/PLACES2014.pdf}}}}
\def\titlerunning{\Title}
\author{
    Janwillem Swalens \qquad
    Stefan Marr \qquad
    Joeri De Koster \qquad
    Tom Van Cutsem
    \institute{
        Software Languages Lab\\
        Vrije Universiteit Brussel, Belgium
    }
    \email{\{jswalens,smarr,jdekoste,tvcutsem\}@vub.ac.be}
}
\def\authorrunning{J. Swalens, S. Marr, J. De Koster, \& T. Van Cutsem}

\PassOptionsToPackage{pdftex,
    colorlinks=false,
    pdfstartview=FitV,
    linkcolor=black,
    citecolor=black,
    urlcolor=black,
    pdfauthor={Janwillem Swalens, Stefan Marr, Joeri De Koster, Tom Van Cutsem},
    pdftitle={\Title},
    pdfsubject={Concurrency Abstractions, Composition, Language Design},
    pdfkeywords={multicore, Clojure, concurrency, STM, actors, programming, CSP, atoms, transactions}}{hyperref}

\newcommand{\citep}[1]{\cite{#1}}
\usepackage{makecell}

\usepackage[inline]{enumitem}

\usepackage{pbox}

\usepackage{subcaption}

\usepackage{color}
\definecolor{check}{RGB}{87,157,28}
\definecolor{cross}{RGB}{197,0,11}
\definecolor{lst-background}{rgb}{0.95,0.95,0.97}
\definecolor{lst-lnumber}{rgb}{0.2,0.2,0.2}
\definecolor{lst-comment}{rgb}{0.1,0.6,0.1}
\definecolor{lst-string}{rgb}{0.3,0.3,0.7}
\definecolor{blocking}{RGB}{191,0,0}
\definecolor{retry}{RGB}{0,102,204}

\usepackage{listings}
\usepackage{amssymb}  % For \circlearrowleft
\usepackage{amsmath}  % For \pmb (bold math symbols)
\newcommand{\cmark}{\textcolor{check}{$\pmb{\checkmark}$}}
\newcommand{\xmark}{\textcolor{cross}{$\pmb{\times}$}}

\newcommand{\eg}{e.\,g.}

\usepackage{comment}
\excludecomment{appendix}  % En/disable appendix

\begin{document}

\clearpage

\maketitle

\begin{abstract}
In the past decades, many different programming models for managing concurrency in applications have been proposed, such as the actor model, Communicating Sequential Processes, and Software Transactional Memory. The ubiquity of multi-core processors has made harnessing concurrency even more important.
We observe that modern languages, such as Scala, Clojure, or F\#, provide not one, but \emph{multiple} concurrency models that help developers manage concurrency.
Large end-user applications are rarely built using just a single concurrency model. Programmers need to manage a responsive UI, deal with file or network I/O, asynchronous workflows, and shared resources.
Different concurrency models facilitate different requirements.
This raises the issue of how these concurrency models interact, and whether they are \emph{composable}.
After all, combining different concurrency models may lead to subtle bugs or inconsistencies.

In this paper, we perform an in-depth study of the concurrency abstractions provided by the Clojure language.
We study all pairwise combinations of the abstractions, noting which ones compose without issues, and which do not.
We make an attempt to abstract from the specifics of Clojure, identifying the general properties of concurrency models that facilitate or hinder composition.
\end{abstract}

\section{Introduction}

% Multi-core
% Concurrency
% models
% combinations

%Since the advent of multi-core processors, concurrency has become unavoidable. To make programming with concurrency easier, a number of concurrency models have been developed. Some programming languages provide multiple concurrency models to the programmer, so that applications with varying requirements can use and combine them. However, the combinations of these models can lead to subtle bugs and inconsistencies in the interactions between the models.

% Expanded / IWT story:

A typical interactive computer program with a graphical user interface needs \emph{concurrency}: a number of activities need to be executed simultaneously. For example, a web browser fetches many files over the network, renders multiple documents in separate tabs, runs plug-ins in the background, and needs to keep the user interface responsive. Since the last decade multi-core processors have become ubiquitous: servers, laptops, and smartphones contain multi-core processors. This has made it possible to perform these concurrent activities simultaneously.

To express the interactions between these activities, a number of concurrency models have been developed. For example, the Actor Model\,\citep{Agha1985} introduces actors to represent components of a system that communicate using asynchronous messages, while Software Transactional Memory (STM)\,\citep{Shavit1995,Harris2005} coordinates the access of concurrent activities to shared memory using transactions.

Because of the varying and extensive requirements of interactive end-user programs, developers choose to combine different models\,\citep{Tasharofi2013}.
For example, a web browser might use actors to run separate tabs in parallel, and manipulate the DOM (Document Object Model) of a web page using STM.
We also see that many programming languages support a number of these models, either through built-in language support or using libraries.
For example, the Akka library\footnote{\url{http://akka.io/}} for Java and Scala provides STM, futures, actors, and agents. Similarly, Clojure\footnote{\url{http://clojure.org/}} has built-in support for atoms, STM, futures, promises, and agents.

However, the concurrency models are not necessarily well-integrated: programmers may experience subtle problems and inconsistencies when multiple concurrency models interact. In this paper, we analyze some problems that arise when combining different concurrency models and outline potential directions for \emph{composable concurrency abstractions}.

\section{Concurrency Models}
\label{sec:models}

% For each model:
% - small definition
% - use case
% - implementation in Clojure

In this paper, we study the combination of atomics, STM, futures and promises, as well as approaches for \emph{communicating threads} (actors and CSP). To illustrate the usefulness
of combining these approaches, we
consider how they could be used in a modern email application.

\vspace{-0.5\baselineskip}
\paragraph{Atomics}
Atomics are variables that support a number of low-level atomic operations, \eg, compare-and-swap.
Compare-and-swap compares the value of an atomic variable with a given value and, only if they are the same, replaces it with a new value.
This is a single atomic operation.
Operations affecting multiple atomic variables are not coordinated, consequently when modifying two atomic variables race conditions can occur.
Atomics are typically used to share independent data fragments that do not require coordinated updates.
For example, a mail client might use an atomic variable to represent the number of unread mails.

In Clojure, \texttt{atom}s are atomic references.
Their value can be read using \texttt{deref}. %and modified using the \texttt{swap!} function.
%Neither operation blocks.
Atoms are modified using \texttt{swap!}, which takes a function that is evaluated with the current value of the atom, and replaces it with the result only if it has not changed concurrently, using compare-and-swap to compare to the original value.
If it did change, the \texttt{swap!} is automatically retried.

\vspace{-0.5\baselineskip}
\paragraph{Software Transactional Memory (STM)} STM\,\citep{Shavit1995,Harris2005} is a concurrency model that allows many concurrent tasks to write to shared memory. Each task accesses the shared memory within a transaction, to manage conflicting operations. If a conflict is detected, a transaction can be retried.
A mail client can use STM to keep information about mails consistent while updates from the server, different devices, and the user are processed at the same time.

Clojure's STM provides \texttt{ref}s, which can only be modified within a transaction. They are read using \texttt{deref} and modified using \texttt{ref-set}. Transactions are represented by a \texttt{dosync} block.
Outside a \texttt{dosync} block, refs can only be read.

\vspace{-0.5\baselineskip}
\paragraph{Futures and Promises} Futures and promises\footnote{Literature uses various different definitions of futures and promises. We use those of amongst others Clojure and Scala here.} are placeholders for values that are only known later.
A future executes a given function in a new thread. Upon completion, the future is resolved to the function's result.
Similarly, a promise is a value placeholder, but it can be created independently and its value is `delivered' later by an explicit operation.
Reading a future or a promise blocks the reading thread until the value is available.
% In general, multiple futures/promises are not coordinated.
A mail client can use futures to render a preview of an attachment in a background thread, while the mail body is shown immediately. When the future completes, the preview can be added to the view.

In Clojure, futures are created using \texttt{future}. Promises are created using \texttt{promise} and resolved using \texttt{deliver}. Futures and promises are read using \texttt{deref}, which potentially blocks.

\vspace{-0.5\baselineskip}
\paragraph{Communicating Threads} We classify concurrency models that use structured communication in terms of messages, instead of relying on shared memory, as \emph{communicating threads}. Often, each thread has only private memory, ensuring that all communication is done via messages. This, combined with having each thread process at most one message at a time, avoids race conditions. However, models and implementations vary the concrete properties to account for a wide range of trade-offs. We distinguish between models that use asynchronous messages, such as actors\,\citep{Agha1985} or agents (as in Clojure and Akka), and models that communicate using synchronous messages, such as CSP\,\citep{Hoare1978}.
In a mail client, typical use cases include the event loop of the user interface
as well as the communication with external systems with strict communication protocols, such as mail servers.

Clojure \texttt{agent}s represent state that can be changed via asynchronous messages. 
\texttt{send} sends a message that contains a function which takes the current state of the agent and returns a new state.
The current state of an agent can be read using a non-blocking \texttt{deref}.
The \texttt{await} function can be used to block until all messages sent to the agent so far have been processed.
Clojure's core.async library implements the CSP model.
A new thread is started using a \texttt{go} block, channels are created using \texttt{chan}.
Inside a \texttt{go} block, values are put on a channel using \texttt{>!} and taken using \texttt{<!}.
Outside \texttt{go} blocks, \texttt{>!!} and \texttt{<!!} can be used.
These operations block until their complementary operation is executed on another thread.

\vspace{0.5\baselineskip}

All operations described above for Clojure are summarized in figure \ref{fig:operations}, highlighting blocking and re-execution.

\begin{figure}
    \small
	\centering
	\newcommand{\blocking}[1]{\textcolor{blocking}{#1 $\otimes$}}
	\newcommand{\retry}[1]{\textcolor{retry}{#1 $\circlearrowleft$}}
	% Less space between rows
	\renewcommand{\arraystretch}{1.0}
	\begin{tabular}{r|cccccc}
		       & atoms  & agents & STM     & futures & promises & core.async\\
		\hline
		create & atom   & agent  & ref     & future  & promise  & chan\\
		read   & deref  & deref  & deref   & \blocking{deref} & \blocking{deref} & \blocking{\textless!}\\
		write  & reset! &        & ref-set &         & deliver  & \blocking{\textgreater!}\\
		       & \retry{swap!} & send   & alter   &         &          & \\
		block  &        &        & \retry{dosync} &         &          & go\\
		other  &        & \blocking{await} &         &         &          & \blocking{\textless!! \textgreater!!}\\
		       &        &        &         &         &          & take! put!\\
	\end{tabular}
	%\vspace{-1ex}
	\caption{Operations supported by Clojure's concurrency models. \blocking{} indicates a (potentially) blocking operation, \retry{} an operation that might be re-executed automatically.}
	\label{fig:operations}
\end{figure}

\section{Integration Problems of Concurrency Models}

\subsection{Criteria for composability}

We study pairwise combinations of the five concurrency models described in the previous section.
Two correctness properties are evaluated: safety and liveness\,\citep{Lamport1977}.
For each combination we study whether additional safety or liveness issues can arise, emerging from the interactions between the two models.
We consider two models \emph{composable} if combining them cannot produce new safety or liveness issues.

\vspace{-0.5\baselineskip}
\paragraph{Safety} Safety means that, given a correct input, a program will not produce an incorrect result.
In our context, many concurrency models are designed to avoid race conditions to achieve safety.
They do this by managing shared resources: STM only allows shared memory to be accessed through transactions, while CSP or the actor model only allow threads to share data through explicit message passing.

When combining two models, new races could be introduced unexpectedly.
For example, some implementations of STM have been proven linearizable\,\citep{Shavit1995}: every concurrent execution is equivalent to a legal sequential execution.
However, this assumes that all shared resources are managed by the STM system, which is not true if a thread communicates with other threads, \eg, using CSP.
This can cause unexpected interleavings that eventually lead to race conditions.
This is not dissimilar to the problem of feature interaction\,\citep{Calder2003}, where several features that each function correctly separately, might behave incorrectly when combined.

Concretely, we study whether the combination of two concurrency models can introduce race conditions: incorrect results caused by unexpected interleavings.

\vspace{-0.5\baselineskip}
\paragraph{Liveness} Liveness guarantees a program will terminate if its input is correct.
In our context, two problems can occur: deadlocks and livelocks.
Deadlocks are introduced by operations that block, waiting until a certain condition is satisfied, but the condition is continually not satisfied.
Livelocks appear when code is re-executed under a certain condition, and this condition is continually satisfied.
Some concurrency models have proven liveness properties, for instance, STMs are usually non-blocking\,\citep{Shavit1995}.
Others try to confine the problem by limiting blocking to a small set of operations.
For example, futures only provide one blocking operation, reading, which waits until the future is resolved. As long as the future eventually resolves, no deadlocks will happen.

Again, when concurrency models are combined, unexpected deadlocks and livelocks might arise.
An STM transaction that uses blocking operations of another model, such as CSP, is not guaranteed to be non-blocking anymore.
Or, a future that contains blocking operations from another model might not ever be resolved, in other words, combining futures with another model can introduce unexpected deadlocks.

We study whether the combination of two concurrency models can introduce new deadlocks, by studying the blocking operations offered by a model, and/or new livelocks, by studying the operations that can cause re-execution.

\subsection{Integration Problems in Clojure}

We examine these combination issues specifically for Clojure.
Each of the five concurrency models from section \ref{sec:models} is embedded in each of the models (including itself).
The complete set of results is shown in the table in figure \ref{fig:tables}.
For example, we embed a \texttt{send} to an agent
\begin{enumerate*}[label=(\arabic*)]
	\item in an atom's \texttt{swap!} block (figure \ref{fig:send:atom}),
	\item in another agent action,
	\item in an STM transaction (figure \ref{fig:send:stm}),
	\item in a future, and
	\item in a \texttt{go} block
\end{enumerate*}
(these results form the second column of figure \ref{fig:tables}).
Even though some of these individual results are already known, we systematically study each pairwise combination of models, in an attempt to provide a comprehensive overview of safety and liveness issues.
A discussion of the most interesting results is given below, a complete discussion of all results in the table is given in the online appendix.

\begin{figure}
    %\small
	\centering
	% Less space between columns
	\setlength{\tabcolsep}{4.8pt}
	% More space between rows
	\renewcommand{\arraystretch}{1.15}
	\begin{tabular}{r|ccccc||ccccc}
		& \multicolumn{5}{c||}{Safety} & \multicolumn{5}{c}{Liveness}\\
		\diaghead{in ~~~~~ used}{in}{used} & atoms & agents & refs & \pbox{20cm}{futures\\[-0.6ex]promises} & channels & atoms & agents & refs & \pbox{20cm}{futures\\[-0.6ex]promises} & channels\\
		\hline
		atoms & \xmark & \xmark & \xmark & \xmark & \xmark & \cmark & \cmark & \cmark & \cmark & \xmark\\
		agents & \cmark & \cmark & \cmark & \cmark & \cmark & \cmark & \cmark & \cmark & \xmark & \xmark\\
		refs & \xmark & \cmark & \cmark & \xmark & \xmark & \cmark & \cmark & \cmark & \cmark & \xmark\\
		\pbox[c]{20cm}{\hfill futures\\[-0.6ex]promises} & \cmark & \cmark & \cmark & \cmark & \cmark & \cmark & \xmark & \cmark & \xmark & \xmark\\
		channels & \cmark & \cmark & \cmark & \cmark & \cmark & \cmark & \xmark & \cmark & \cmark & \xmark\\
	\end{tabular}
	\caption{This table shows when safety and liveness issues can arise by combining two models in Clojure. The model in the column is used in the model in the row.}
	\label{fig:tables}
\end{figure}

\begin{figure}
	\centering
	\begin{subfigure}[t]{0.48\textwidth}
		\centering
		\begin{lstlisting}
(def notifications (agent '()))
(def unread-mails (atom 0))

(swap! unread-mails
  (fn [n]
    (send notifications
      (fn [msgs] (cons "New mail!" msgs)))
    (inc n)))
		\end{lstlisting}
		\vspace{-1ex}
		\subcaption{Sending a message to an agent, in \texttt{swap!}. Send may happen more than once.}
		\label{fig:send:atom}
	\end{subfigure}\hfill
	\begin{subfigure}[t]{0.5\textwidth}
		\centering
		\begin{lstlisting}
(def notifications (agent '()))
(def mail (ref {:subject "Hi" :archived false}))

(dosync
  (ref-set mail (assoc @mail :archived true))
  (send notifications
    (fn [msgs] (cons (str "Archived mail " (:subject @mail)) msgs))))
		\end{lstlisting}
		\vspace{-1ex}
		\subcaption{Sending a message to an agent, in a \texttt{dosync}. Send is delayed until the transaction is committed.}
		\label{fig:send:stm}
	\end{subfigure}
	%\vspace{-1ex}
	\caption{Sending a message to an agent, in a block that might re-execute (\texttt{swap!} and \texttt{dosync}).}
	\label{fig:send}
\end{figure}

\vspace{-0.5\baselineskip}
\paragraph{Safety}
We first look at the possibility of race conditions (left side of table \ref{fig:tables}).
Race conditions are caused by an incorrect interleaving between two models.

When using any concurrency model in the function given to an atom's \texttt{swap!} operation (first row of the table), race conditions are possible, because the function might be re-executed if the atom changed concurrently. For example, when this function sends a message to an agent, it could be sent twice (figure \ref{fig:send:atom}). Moreover, because operations on multiple atoms are not coordinated, their updates are inherently racy.

For STM, actions inside a \texttt{dosync} block are re-executed if the transaction is retried, and therefore no irrevocable operations should happen inside \texttt{dosync}. However, there are two safe combinations. Firstly, when a message is sent to an agent inside a \texttt{dosync} block (figure \ref{fig:send:stm}), Clojure does not send this message immediately. Instead, it delays the send until the transaction is successfully committed. Secondly, embedding one \texttt{dosync} block in another means the inner transaction will be merged with the outer one, and as a result transactions are combined safely.

Based on these results we conclude that
\emph{if the `outer' model might re-execute code it is given,
and the `inner' model might perform irrevocable actions,
unexpected interleavings can happen and therefore
safety is not guaranteed}.

\vspace{-0.5\baselineskip}
\paragraph{Liveness}
Next, we look at the liveness property (right side of table \ref{fig:tables}): deadlocks and livelocks.

\emph{Deadlocks} are introduced by blocking operations (indicated in figure \ref{fig:operations}). CSP relies heavily on blocking for communication, and as such deadlocks are possible when it is embedded in another model (see last column). This is particularly problematic when embedded in \texttt{swap!} (atoms) or \texttt{dosync} (STM): synchronous communication is irrevocable and should not be re-executed.

Reading a future or a promise blocks until its value is available. This can cause a deadlock when a promise is embedded in an agent (fourth column), because one thread might send an action to an agent, which reads a promise that is delivered in a later action sent to that same agent, possibly from another thread.
Reading futures inside another future can also cause a deadlock when mutually recursive futures are allowed, as is the case in Clojure.

Finally, the agents' blocking \texttt{await} operation can cause deadlocks in a \texttt{go} block or a future (second column). In agent actions and STM transactions this situation is prevented by raising an exception.
We conclude that
\emph{if the `inner' model might block,
and the `outer' model does not expect this,
a deadlock is possible}.

\emph{Livelocks} appear when code is re-executed (operations that can cause re-execution are indicated in figure \ref{fig:operations}).
An STM transaction is retried when it conflicts with another one, causing a livelock if the conflict would consistently occur. However, Clojure's STM prevents such deadlocks and livelocks dynamically\,\citep{emerick2011clojure}.
In general, \emph{a livelock can appear when a model that re-executes code is combined with a model that causes this re-execution to continually happen}. However, this occurs in none of the examined cases in Clojure.

\section{Solutions and Open Questions}

In the discussion of the previous section, we already pointed out some places where bad interactions between models are avoided by Clojure. Specifically:
\begin{enumerate*}[label=(\arabic*)]
	\item Sending a message to an agent in an STM transaction is delayed until the transaction has been committed.
	\item \texttt{await} is not allowed in STM transactions, nor in actions sent to agents.
	\item A \texttt{dosync} block embedded in another will not start a new transaction: the inner transaction is merged into the outer one.
%	\item (Delivering a promise more than once does not raise an error, but instead fails silently.) \jw{Not discussed before}
\end{enumerate*}
These mechanisms could be replicated in some other cases: similar to sending a message to an agent in a transaction, delivering a promise could also be delayed until the transaction has been committed. The combination of futures and transactions could further be improved by canceling futures started in a transaction if the transaction is aborted.

A solution to deadlocks caused by the combination of agents and futures/promises is to disallow reading a future/promise in an agent action, an operation that potentially blocks the agent. Instead, the future/promise would need to be read before sending the message.

To remove the unsafe combinations of atoms with any other model, some of the mechanisms of combinations with STM could be replicated (e.\,g., delaying sends to agents). However, this can be considered contrary to the purpose of atoms: they are low-level and uncoordinated, and accordingly offer no safety guarantees.
Similarly, the CSP model uses synchronous, blocking operations by design, and therefore liveness issues are inherent to this model. Trying to avoid these would be contradictory to the nature of the model.

In general, in future research we would like to decompose several concurrency models into their components, or ``building blocks''.
For example, we observe that some common elements are:
\begin{enumerate*}[label=(\arabic*)]
	\item most models supply a way to create new threads (e.\,g., \texttt{agent}, \texttt{future}, \texttt{go});
	\item agents, actors, futures, promises and CSP provide message passing (asynchronous or synchronous);
	\item agents, actors, and CSP have private memory per thread, while
	\item atomics and STM provide a way to manage shared memory.
\end{enumerate*}

We want to extract such common elements and provide a way to compose them safely and efficiently.
For example, different threads could have some private memory, communicate using message passing (as in the actor model and CSP), but also share a section of memory (\eg, using STM).
Using these composable concurrency abstractions, it should be possible to express existing concurrency models as well as combinations of them correctly.
In the end, it should be possible to write complex applications, such as the email client example of section \ref{sec:models}, using a combination of concurrency models, without introducing new safety or liveness issues caused by interactions between the models.

\section{Conclusion and Future Work}

% 1. Many concurrency models, they are combined (in languages and in actual programs)
% 2. Study problems that can appear when combining, specifically Clojure. Identified 4 reasons for conflicts.
% 3. Look at solutions, some already there, some new ones.
% 4. Future work: work towards composable concurrency abstractions: decompose existing models into "building blocks", provide a way to compose these safely and efficiently.

There exist various different concurrency models, and in many large-scale applications these are combined. However, subtle problems and inconsistencies can appear in the interactions between these models. In this paper, we studied the safety and liveness issues that can appear when the various concurrency models available for Clojure are combined.

We identified four reasons for conflicts between models.
Firstly, when a model re-executes code, and this code uses another concurrency model that performs irrevocable actions, safety is not guaranteed. Clojure takes some special precautions in some of these cases.
Secondly, when a model re-executes code, and this code can cause the re-execution to continually happen, a livelock is possible. In Clojure, this is prevented in the studied cases.
Thirdly, when a model that supports blocking operations is embedded in a model that does not expect this, deadlocks become possible. Again, Clojure prevents this in some cases but not in others.
Lastly, some models do not provide safety or liveness guarantees by design.

In future work, we aim to work towards composable concurrency abstractions: we will decompose existing concurrency models into more primitive ``building blocks'', and provide a way to compose these safely and efficiently.

\bibliographystyle{eptcs}
\bibliography{bibliography}

\begin{appendix}

\newpage
\section{Appendix}

This appendix discusses table \ref{fig:tables} in more detail---not only unsafe but also safe combinations. We first look at the safety property, next at liveness.

\subsection{Safety}

The following sections describe table \ref{fig:safe} \emph{row by row}: for each model we examine whether another model can be safely embedded in it.

\begin{figure}[h]
	\centering
	% Less space between columns
	%\setlength{\tabcolsep}{3pt}
	% More space between rows
	\renewcommand{\arraystretch}{1.2}
	\begin{tabular}{r|ccccc}
		\diaghead{in ~ used}{in}{used}& atoms & agents & refs & \pbox{20cm}{futures\\[-0.6ex]promises} & channels\\
		\hline
		atoms    & \xmark & \xmark & \xmark & \xmark & \xmark\\
		agents   & \cmark & \cmark & \cmark & \cmark & \cmark\\
		refs     & \xmark & \cmark & \cmark & \xmark & \xmark\\
		\pbox[c]{20cm}{\hfill futures\\[-0.6ex]promises} & \cmark & \cmark & \cmark & \cmark & \cmark\\
		channels & \cmark & \cmark & \cmark & \cmark & \cmark\\
	\end{tabular}
	\caption{This table shows when safety issues (race conditions) can arise by combining two models. The model in the column is used in the model in the row.}
	\label{fig:safe}
\end{figure}

\subsubsection{... Used in swap! (Atoms)}

\begin{figure}[t]
	\centering
	\begin{subfigure}[t]{0.45\textwidth}
		\centering
		\vspace{11pt}
		\begin{lstlisting}
; Number of read and unread mails
(def unread (atom 10))
(def read (atom 20))
; Thread 1: read all mail
(swap! read (fn [n] (+ n @unread)))
(reset! unread 0)
; Thread 2: mark mail as unread
(swap! unread inc)
(swap! read dec)
		\end{lstlisting}
		\subcaption{Atoms are uncoordinated, and therefore race conditions are possible. When this code is executed using two threads, the sum of unread and read mails can change erroneously.\\}
		\label{fig:safe:atom:atom}
	\end{subfigure}\hfill
	\begin{subfigure}[t]{0.53\textwidth}
		\centering
		\begin{lstlisting}
(def unread (atom 10))
(def mail (ref {:subject "Hi" :read-count 0}))

(swap! unread
  (fn [n]
    (dosync
      (ref-set mail
        (assoc @mail :read-count
          (inc (:read-count @mail)))))
    (dec n)))
		\end{lstlisting}
		\subcaption{When a mail is read, we decrement the ``unread mail'' counter (shown in the UI) and increment the ``read-count'' of the mail. The \texttt{dosync} block cannot only be retried (safe), but also be re-executed when the function passed to \texttt{swap!} is re-executed (unsafe).}
		\label{fig:safe:atom:ref}
	\end{subfigure}
	\caption{Combinations with atoms.}
\end{figure}

In general, combinations with atoms are unsafe: the function given to \texttt{swap!} might be executed more than once, and therefore cannot contain irrevocable actions.

\begin{description}[leftmargin=0.5cm,itemsep=0ex]
	\item[Atoms] Combining atoms, as illustrated in figure \ref{fig:safe:atom:atom}, can lead to race conditions, because there is no coordination between atoms.
	\item[Agents] A \texttt{send} in a \texttt{swap!} block could be repeated, as shown in figure \ref{fig:send:atom}: this is unsafe.
	\item[Refs] A \texttt{dosync} block in a \texttt{swap!} block could be executed multiple times, which is unsafe. This is illustrated in figure \ref{fig:safe:atom:ref}.
	\item[Futures/promises] Reading a future or a promise in a \texttt{swap!} is safe: it can increase the chance of the code being re-executed, but does not introduce race conditions. Creating a future or a promise in a \texttt{swap!} block can cause it to be created multiple times, which is undesirable but safe. Delivering a promise in a \texttt{swap!} block is not safe however: only the first deliver will succeed, subsequent ones will fail silently\footnote{Delivering a promise more than once will fail silently in Clojure 1.3 or later, where subsequent calls to \texttt{deliver} return \texttt{nil}. In previous versions, subsequent delivers would throw an exception.}.
	\item[Channels] Starting a \texttt{go} block and reading from or writing to channels can be repeated in a \texttt{swap!}, this is unsafe.
\end{description}

\subsubsection{... Used in Agent Actions}

Agents are given an action asynchronously using \texttt{send}, which is scheduled to be executed in the agent thread. As only one action will be active per agent, and this action will only be executed once, there are generally no safety problems in using other concurrency models in agent actions.

\begin{description}[leftmargin=0.5cm,itemsep=0ex]
	\item[Atoms] Both reading and changing atoms in agent actions is safe, and common in Clojure code.
	\item[Agents] It is possible to send actions to other agents, or the same agent, in an agent action, this is an asynchronous operation and causes no problems.
	\item[Refs] Executing a \texttt{dosync} block in an agent action is safe.
	\item[Futures/promises] Reading a future or promise in an agent action can block, but is safe. Creating a future or promise is safe as well. Delivering a promise is a common way for an agent to return a value.
	\item[Channels] Starting a \texttt{go} block in an agent action is safe (another thread is started), reading from and writing to channels as well.
\end{description}

\subsubsection{... Used in Transactions (STM)}
\label{appendix:safe:refs}

As a \texttt{dosync} block might be executed more than once, it is generally unsafe to combine other models with STM. The implementation of Clojure foresees two exceptions: using agents in a transaction, and embedding one \texttt{dosync} block in another.

\begin{description}[leftmargin=0.5cm,itemsep=0ex]
	\item[Atoms] Modifying atoms in a transaction is unsafe: the modification can be executed more than once.
	\item[Agents] Clojure deals with message sends to agents in a \texttt{dosync} block in a special way: the send is delayed, until the transaction has been successfully committed. Therefore, using agents in a transaction is safe.
	\item[Refs] A \texttt{dosync} block embedded in another is merged into one transaction. Therefore this is a safe combination.
	\item[Futures/promises] Reading a future or a promise in a transaction is safe, although it can increase the chance of the transaction failing on the first execution (on subsequent executions, the read will not block). Creating a future or promise can be undesirable, but is generally safe. However, delivering a promise is unsafe: only the first deliver succeeds, subsequent ones fail silently. A proposed solution is to delay the deliver until the transaction is committed---similar to \texttt{send} (agents).
	\item[Channels] Starting a \texttt{go} block in a transaction, as well as reading from and writing to channels, are irrevocable actions that should not be repeated, and are therefore unsafe.
\end{description}

\subsubsection{... Used in Futures}

Starting a future simply creates another thread. In general, it is safe to use other concurrency models in this new thread.

\begin{description}[leftmargin=0.5cm,itemsep=0ex]
	\item[Atoms] Both reading and modifying atoms in a future is safe.
	\item[Agents] Reading, sending and awaiting agents in a future is safe.
	\item[Refs] Transactions can be executed safely in futures, and commonly are.
	\item[Futures/promises] In a future, it is safe to read another future or promise, to deliver a promise, and to create another future.
	\item[Channels] In a future it is safe to create a \texttt{go} block (this simply starts another thread), as well as to read from and write to channels. In many ways, \texttt{future} and \texttt{go} are similar: both start a new thread, \texttt{future} returns a future that will contain the result of its body when dereferenced while \texttt{go} returns a channel that will receive the result of the body.
\end{description}

\subsubsection{... Used in Go Blocks (CSP)}

\texttt{go} starts a new thread, in which it is generally safe to use other concurrency models.

\begin{description}[leftmargin=0.5cm,itemsep=0ex]
	\item[Atoms] Atoms can be read and modified safely in \texttt{go} blocks.
	\item[Agents] Agents can be read, sent to and awaited safely in \texttt{go} blocks.
	\item[Refs] It is safe to embed a \texttt{dosync} block in a \texttt{go} block. It is not possible to execute \texttt{<!} and \texttt{>!} in a transaction, as these operations can only be used directly in a \texttt{go} block (and not in a \texttt{do} block, such as \texttt{dosync}). Using \texttt{<!!} or \texttt{>!!} in a \texttt{dosync} block is unsafe, as already discussed in section \ref{appendix:safe:refs}.
	\item[Futures/promises] Reading futures and promises, creating them, and delivering promises is safe in a \texttt{go} block.
	\item[Channels] It is safe to start a \texttt{go} block in another \texttt{go} block.
\end{description}

\subsection{Liveness}

The following sections describe table \ref{fig:live} \emph{column by column}: for each model we examine the liveness issues that arise when it is embedded into another model.

\begin{figure}[h]
	\centering
	% Less space between columns
	%\setlength{\tabcolsep}{3pt}
	% More space between rows
	\renewcommand{\arraystretch}{1.2}
	\begin{tabular}{r|ccccc}
		\diaghead{in ~ used}{in}{used}& atoms & agents & refs & \pbox{20cm}{futures\\[-0.6ex]promises} & channels\\
		\hline
		atoms    & \cmark & \cmark & \cmark & \cmark & \xmark\\
		agents   & \cmark & \cmark & \cmark & \xmark & \xmark\\
		refs     & \cmark & \cmark & \cmark & \cmark & \xmark\\
		\pbox[c]{20cm}{\hfill futures\\[-0.6ex]promises} & \cmark & \xmark & \cmark & \xmark & \xmark\\
		channels & \cmark & \xmark & \cmark & \cmark & \xmark\\
	\end{tabular}
	\caption{This table shows when liveness issues (deadlocks or livelocks) can arise by combining two models.}
	\label{fig:live}
\end{figure}

\subsubsection{Atoms Used in ...}

\begin{figure}[t]
	\centering
	\begin{subfigure}[t]{0.53\textwidth}
		\centering
		\vspace{88pt}
		\begin{lstlisting}
(def a (atom 0))
(swap! a
  (fn [x]
    (swap! a inc)))
		\end{lstlisting}
		\subcaption{\texttt{swap!} is called in another \texttt{swap!} of the \emph{same} atom. This code will always lead to a livelock: the outer \texttt{swap!} executes its body in which the inner \texttt{swap!} modifies the atom and therefore causes another execution of the outer \texttt{swap!}.}
		\label{fig:live:atom:atom:same}
	\end{subfigure}\hfill
	\begin{subfigure}[t]{0.45\textwidth}
		\centering
		\begin{lstlisting}
(def a (atom 0))
(def b (atom 0))
; Thread 1
(swap! a
  (fn [x]
    (swap! b inc)
    (inc x))))
; Thread 2
(swap! b
  (fn [x]
    (swap! a inc)
    (inc x))))
		\end{lstlisting}
		\subcaption{Two threads modify two atoms. The functions given to \texttt{swap!} might get re-executed several times, until eventually they happen to not overlap and they succeed.}
		\label{fig:live:atom:atom:diff}
	\end{subfigure}
	\caption{Liveness issues when combining atoms.}
\end{figure}

Atoms do not support any blocking operations, and can therefore not cause any deadlocks. The re-execution of \texttt{swap!} blocks can potentially lead to livelocks however.

\begin{description}[leftmargin=0.5cm,itemsep=0ex]
	\item[swap! (Atoms)] A livelock will always occur when a \texttt{swap!} block is called inside another \texttt{swap!} block on the \emph{same} atom (figure \ref{fig:live:atom:atom:same}). When modifying two different atoms, the program will eventually terminate (figure \ref{fig:live:atom:atom:diff}).
	\item[Agent Actions] No liveness issues arise when using atoms in an agent action.
	\item[Transactions] No issues.
	\item[Futures] No issues.
	\item[Go Blocks] No issues.
\end{description}

\subsubsection{Agents Used in ...}

\begin{figure}[p]
	\centering
	\begin{subfigure}[t]{0.46\textwidth}
		\centering
		\begin{lstlisting}
(def at (atom 0))
(def ag (agent 0))

(swap! at
  (fn [_]
    (send ag
      (fn [_]
        (swap! at inc)))
    (await ag)
    inc))
		\end{lstlisting}
		\subcaption{When a \texttt{swap!} contains a function that sends a message to an agent which modifies the atom, \emph{and} then awaits the agent until the message has been processed, a livelock is possible. This is in fact just an extended version of the example in figure \ref{fig:live:atom:atom:same}.}
		\label{fig:live:agent:atom}
	\end{subfigure}\hfill
	\begin{subfigure}[t]{0.52\textwidth}
		\centering
		\begin{lstlisting}
(def c (chan))
(def ag (agent 0))

; Thread 1
(send ag (fn [_] (<!! c)))

; Thread 2
(go
  (await ag)
  (>! c "test"))
		\end{lstlisting}
		\subcaption{Thread 1 sends a message to the agent, which takes a value from a channel. In the thread that writes to this channel, an \texttt{await} is inserted before the write. This can lead to a deadlock. In complex programs, channels and agents can be passed between, read from and written to by different threads and such mistakes could be much more subtle.}
		\label{fig:live:agent:csp}
	\end{subfigure}\hfill
	\begin{subfigure}[t]{\textwidth}
		\centering
		\begin{lstlisting}
(def mail-ui (agent {:subject "Hi" :thumbnails []}))

(defn generate-thumbnail [attachment]
  (future
    (let [thumbnail (create-thumbnail attachment)]
      (if (nil? thumbnail)
        false
        (do
          (send mail-ui
            (fn [m] (assoc m :thumbnails (conj (:thumbnails m) thumbnail))))
          (await mail-ui) ; Make sure UI has updated before proceeding
          true))

(let [thumbnail1 (generate-thumbnail attachment1)]
  (send mail-ui
    (fn [m]
      ; Modify the UI based on whether thumbnail could be generated or not
      (if @thumbnail1
        ...
        ...)))
		\end{lstlisting}
		\subcaption{This code sample generates thumbnails for e-mail attachments. \texttt{create-thumbnail} is a computationally intensive function that creates a thumbnail given an attachment, or returns \texttt{nil} if no thumbnail could be generated. We use a future to offload the thumbnail generation to another thread, this future will resolve to \texttt{true} or \texttt{false} depending on whether the thumbnail could be generated. An agent represents the state of the UI. If the second \texttt{send} occurs before the \texttt{await} in the future, this code will deadlock: the agent action is waiting for the future to finish (\texttt{@thumbnail1}) while the future is waiting for the agent to finish (\texttt{await}).}
		\label{fig:live:agent:future}
	\end{subfigure}
	\caption{Liveness issues when combining agents.}
\end{figure}

Agents support \texttt{await}, a function that blocks until all messages sent to that agent up to that point in time have been processed. This can cause a deadlock in case the messages block waiting for an action that happens after the \texttt{await}.

\begin{description}[leftmargin=0.5cm,itemsep=0ex]
	\item[swap! (Atoms)] Awaiting an agent in a \texttt{swap!} block increases the chance of the \texttt{swap!} failing and re-executing. If the message sent to the agent causes the \texttt{swap!} to fail, a livelock is possible. This is shown in figure \ref{fig:live:agent:atom}.
	\item[Agent Actions] Possible deadlocks are prevented by Clojure, as it does not allow \texttt{await} in actions send to agents.
	\item[Transactions] Issues are again prevented in Clojure by disallowing \texttt{await} in \texttt{dosync}.
	\item[Futures] A deadlock can arise when an agent is \texttt{await}ed in a future, and that future is dereferenced in another action sent to that agent (illustrated in figure \ref{fig:live:agent:future}).
	\item[Go Blocks] Deadlocks are possible when \texttt{await} and CSP's blocking operations on channels are combined (e.\,g., figure \ref{fig:live:agent:csp}).
\end{description}

\subsubsection{Refs Used in ...}

In general, STM does not provide any blocking operations and therefore cannot introduce deadlocks. A livelock occurs if a transaction is continually retried, whether this is prevented depends on the STM implementation.

\begin{description}[leftmargin=0.5cm,itemsep=0ex]
	\item[swap! (Atoms)] No liveness issues arise when embedding a \texttt{dosync} in a \texttt{swap!}.
	\item[Agent Actions] There are no issues when embedding a transaction in an agent action.
	\item[Transactions] Livelocks could exist if a transaction would be continually retried, however, this is dynamically prevented by the STM implementation of Clojure\,\citep{emerick2011clojure}.
	\item[Futures] \texttt{dosync} can be used in a future without issues.
	\item[Go Blocks] \texttt{dosync} can be used in \texttt{go} blocks, however \texttt{<!!} and \texttt{>!!} can cause problems when used in the \texttt{dosync} block (covered in section \ref{appendix:live:csp}).
\end{description}

\subsubsection{Futures/promises Used in ...}

\begin{figure}[b]
	\centering
	\begin{subfigure}[t]{0.65\textwidth}
		\centering
		\begin{lstlisting}
(def p (promise))
(def ag (agent 0))
(send ag (fn [_] @p))
(send ag (fn [_] (deliver p 1)))
		\end{lstlisting}
		\subcaption{This program will deadlock, as the agent blocks until the promise is resolved, but this promise will only be resolved by that same agent in a later action sent to the same agent. This example is simple, in more complex programs the two \texttt{send}s might come from different threads and contain more complex logic.}
		\label{fig:live:future:agent}
	\end{subfigure}\hfill
	\begin{subfigure}[t]{0.33\textwidth}
		\centering
		\begin{lstlisting}
(declare f2)
(def f1 (future (f @f2)))
(def f2 (future (g @f1)))
		\end{lstlisting}
		\vspace{11pt}
		\subcaption{This program always deadlocks. The programmer must not write mutually recursive futures.}
		\label{fig:live:future:future}
	\end{subfigure}
	\caption{Liveness issues when combining futures.}
\end{figure}

Liveness issues arise when a future or promise is read, but never resolved. In some cases, mistakes by the programmer can cause such a situation.

\begin{description}[leftmargin=0.5cm,itemsep=0ex]
	\item[swap! (Atoms)] Reading a future or a promise in a \texttt{swap!} can block, but won't block again if the \texttt{swap!} is re-executed. Therefore no new liveness issues arise.
	\item[Agent Actions] A deadlock can occur when a promise is read by an agent, before it would resolved by the same agent. This is illustrated in figure \ref{fig:live:future:agent}.
	\item[Transactions] Reading a future/promise in a transaction can block, but if the transaction is re-executed the read will not block anymore. No new liveness issues arise.
	\item[Futures] Clojure allows mutually recursive futures, i.\,e., one future could be waiting on another while the second future is waiting on the first. This is illustrated in figure \ref{fig:live:future:future}.
	\item[Go Blocks] Reading a future/promise can block the go routine, however, this poses no problem.
\end{description}

\subsubsection{Channels Used in ...}
\label{appendix:live:csp}

The CSP model uses blocking operations, \texttt{<!} and \texttt{>!}, to read from and write to channels. As such, deadlocks are always possible, even without combining the model with others. Additional issues can arise when these (irrevocable) operations are used in code that might get re-executed (\texttt{swap!} for atoms and \texttt{dosync} for STM), as the thread with which they are communicating has already proceeded.

\begin{description}[leftmargin=0.5cm,itemsep=0ex]
	\item[swap! (Atoms)] Starting \texttt{go} blocks, as well as reading from and writing to channels, are irrevocable actions, using them in a \texttt{swap!} block can lead to deadlocks when the \texttt{swap!} is re-executed.
	\item[Agent Actions] If \texttt{<!!} or \texttt{>!!} block in an agent, the agent is not guaranteed to progress.
	\item[Transactions] Starting \texttt{go} blocks, as well as reading from and writing to channels, are irrevocable actions and should not be used in a transaction as the transaction might be retried.
	\item[Futures] Using \texttt{<!!} and \texttt{>!!} in a future is similar to using \texttt{<!} and \texttt{>!} in a \texttt{go} block. No additional liveness issues arise except those inherent to the CSP model.
	\item[Go Blocks] The programmer can make mistakes, e.\,g., reading from a channel but forgetting to write something to it, leading to deadlocks.
\end{description}

\end{appendix}

\end{document}